\documentclass{Interspeech}

\interspeechcameraready

\usepackage{amsmath}
\newcommand\norm[1]{\lVert#1\rVert}

\usepackage{multirow} %
\usepackage{algorithm}
\usepackage{algpseudocode}
\usepackage{svg}
\usepackage{comment}
\usepackage{mathtools}
\usepackage{xfrac}

\newcommand{\D}[1]{\mathrm{d}{#1}}

\usepackage{esvect}
\usepackage{adjustbox}
\usepackage{multirow}
\newcommand{\seqt}{\overrightarrow{t}}
\newcommand{\Vt}{\mathbf V^{\overrightarrow{t}}}
\newcommand{\Vtfk}{V^{\overrightarrow{t}}(l,k)}

\DeclareMathOperator*{\argmin}{arg\,min}

\usepackage[font=small,skip=0pt]{caption}
\usepackage[nolist, nohyperlinks]{acronym}

\usepackage{makecell} %

\begin{acronym}
\acro{bb}[BB]{Brownian bridge}
\acro{db}[DB]{Diffusion Buffer}
\acro{sgm}[SGM]{score-based generative model}
\acro{snr}[SNR]{signal-to-noise ratio}
\acro{gan}[GAN]{generative adversarial network}
\acro{vae}[VAE]{variational autoencoder}
\acro{ddpm}[DDPM]{denoising diffusion probabilistic model}
\acro{stft}[STFT]{short-time Fourier transform}
\acro{istft}[iSTFT]{inverse short-time Fourier transform}
\acro{sde}[SDE]{stochastic differential equation}
\acro{ode}[ODE]{ordinary differential equation}
\acro{ou}[OU]{Ornstein-Uhlenbeck}
\acro{ve}[VE]{Variance Exploding}
\acro{dnn}[DNN]{deep neural network}
\acro{pesq}[PESQ]{Perceptual Evaluation of Speech Quality}
\acro{se}[SE]{speech enhancement}
\acro{tf}[T-F]{time-frequency}
\acro{elbo}[ELBO]{evidence lower bound}
\acro{WPE}{weighted prediction error}
\acro{PSD}{power spectral density}
\acro{RIR}{room impulse response}
\acro{SNR}{signal-to-noise ratio}
\acro{LSTM}{long short-term memory}
\acro{POLQA}{Perceptual Objectve Listening Quality Analysis}
\acro{SDR}{signal-to-distortion ratio}
\acro{ESTOI}{Extended Short-Term Objective Intelligibility}
\acro{ELR}{early-to-late reverberation ratio}
\acro{TCN}{temporal convolutional network}
\acro{DRR}{direct-to-reverberant ratio}
\acro{nfe}[NFE]{number of function evaluations}
\acro{rtf}[RTF]{real-time factor}
\end{acronym}

\title{Diffusion Buffer: Online Diffusion-based Speech Enhancement with Sub-Second Latency}
\author[affiliation={1, 2}]{Bunlong}{Lay}
\author[affiliation={1}]{Rostislav}{Makarov}
\author[affiliation={1}]{Timo}{Gerkmann}

\affiliation{University of Hamburg}{Signal Processing}{Germany}
\affiliation{}{Hamburger Informatik Technologie-Center e.V.}{Germany}
\email{bunlong.lay@uni-hamburg.de, rostislav.makarov@uni-hamburg.de, timo.gerkmann@uni-hamburg.de}

\usepackage{comment}

\begin{document}

\maketitle
\begin{abstract}
Diffusion models are a class of generative models that have been recently used for speech enhancement with remarkable success but are computationally expensive at inference time. Therefore, these models are impractical for processing streaming data in real-time. In this work, we adapt a sliding window diffusion framework to the speech enhancement task. Our approach progressively corrupts speech signals through time, assigning more noise to frames close to the present in a buffer. This approach outputs denoised frames with a delay proportional to the chosen buffer size, enabling a trade-off between performance and latency. Empirical results demonstrate that our method outperforms standard diffusion models and runs efficiently on a GPU, achieving an input-output latency in the order of 0.3 to 1 seconds. This marks the first practical diffusion-based solution for online speech enhancement\footnote{The source code is publicly available at \\\url{https://github.com/sp-uhh/Diffusion-Buffer}}.
\end{abstract}

\section{Introduction} \label{sec:intro}

The objective of speech enhancement (SE) is to retrieve the original clean speech signal from a noisy mixture that is affected by additive environmental noise ~\cite{hendriks2013dft}. Online SE, which refers to enhancing the speech signal with a limited latency as it is being received, holds significant importance in a variety of applications. The ability to perform SE in an online fashion is critical for ensuring clear and intelligible communication in video conferencing, VoIP calls, and other live interaction platforms. The specific latency constraints depend on the use case. In some cases, such as in live captioning for broadcast media or online streaming, a relatively large latency of 0.1 up to 1 second is also acceptable.
However, developing online SE systems is challenging. It requires low processing time under given hardware requirements which makes the use of computationally costly methods impractical. Therefore, online SE has to use a relatively small model that is capable of handling real-world data unseen during the development of these systems.

Traditional SE approaches attempt to leverage statistical relationships between the clean speech signal and the surrounding environmental noise  ~\cite{gerkmann2018book_chapter}. In recent years, various machine learning techniques have been introduced, treating SE as a discriminative learning task \cite{wang2018supervised, luo2019conv, semamba}. In contrast to these approaches that learn a direct mapping from a corrupted input file to clean speech, so-called \emph{diffusion models} transform a known tractable prior distribution to a specific data distribution, which is in general intractable.
The fundamental concept behind these generative models involves iteratively adding Gaussian noise to the data \emph{forward process}, thereby transforming data into a tractable distribution such as the Gaussian distribution. A neural network (NN) is trained to invert this diffusion process as part of a so-called \emph{reverse process}~\cite{ho2020denoising}. In the context of SE, the prior distribution is the distribution of mixture signals corrupted by Gaussian noise. Hence, solving the reverse process means enhancing the corrupted mixtures as the reverse process transforms the distribution of corrupted mixtures into the distribution of clean speech signals. Diffusion models have proven to achieve excellent performance results when tested on unseen data and different corruption types \cite{journal, scheibler24universe, Nortier2023UnsupervisedSE}. This ability makes diffusion models interesting for real-world scenarios where audio data is streamed live.

However, the reverse process typically employs a large NN that is called several times. Therefore, current diffusion-based models for SE are too slow for the task of online processing of streamed audio data. Recent advancements in diffusion-based approaches \cite{rollingdiff, fifo}, have demonstrated the ability to generate videos by progressively denoising temporal data over time. Inspired by these methods, we propose adapting the concept of temporal denoising for diffusion-based SE. To achieve this, we introduce a \ac{db} containing the most recent $B$ frames of streamed noisy data. The latest input frame is always placed at the end of this buffer. 
Within this buffer, frames closer to the present time remain noisier, while frames further in the past (towards the beginning of the buffer) are progressively denoised. A frame that reaches the very beginng of the buffer is removed from the \ac{db} and directed to the ouput, making room for a new noisy frame from the stream. 
We show that by adjusting the buffer size we can trade enhancement performance for latency. Moreover, we demonstrate that in terms of PESQ and WVMOS we perform even slightly better than vanilla score-based diffusion models \cite{lay202interspeech, journal} when we take 60 reverse steps. The proposed method achieves comparable results to the offline case when the latency introduced by the buffer is set to 320--960 ms. A key advantage of this method is that the score model is only called once per input frame. In fact, with the proposed method streamed data is processed in an online-fashion on a laptop with an NVIDIA RTX 4080 GPU with algorithmic latencies of 320 - 960~ms. The real-time factor (RTF) is below $1$ and thus the input-output latencies are approximately given by the algorithmic latency plus hop size.

\section{Background}
We consider a speech enhancement task for speech signals corrupted by additive noise. The input is a noisy mixture $\mathbf Y = \mathbf S+\mathbf E$ in the complex \ac{stft} domain, consisting of a clean speech signal $\mathbf S \in \mathbb{C}^{F \times K}$ and environmental noise $\mathbf E \in \mathbb{C}^{F \times K}$, where $K$ and $F$ are the number of frames and number of frequency bins in $\mathbf Y$, respectively. The output is an estimate $\hat{\mathbf S}$ of the clean speech signal. While bold symbols (e.g. $\mathbf Y$) refer to spectrograms, we denote a single coefficient of a time-frequency bin by $Y(l,k) \in \mathbb{C}$ for $1\leq l \leq F$ and $1\leq k \leq K$.

\subsection{Stochastic Differential Equations} \label{sec:sde}
The following formulation applies simultaneously to all time-frequency bins of $Y(l,k)$. This means that the following equations are one-dimensional. For readability, we will omit indices $l,k$ in this subsection and reintroduce them in the following sections whenever the indices are important. Following the approach in \cite{lay202interspeech, journal}, we model the forward process of the score-based generative model with a \acf{sde} defined on $0 \leq t < T_{\text{max}}$:
\begin{equation} \label{eq:fsde}
    \D{X_t} =
       f(X_t, Y) \D{t}
        + g(t)\D{{ w}},
\end{equation}
where $w$ is the standard Wiener process \cite{kara_and_shreve}, $X_t$ is the current process state with initial condition $X_0 =  S$, and $t$ is a continuous diffusion time-step variable describing the progress of the process ending at the last diffusion time-step $T_{\text{max}}$.
The \emph{drift coefficient} $f(X_t, Y) \D{t}$ can be integrated by Lebesgue integration \cite{rudin}, and the \emph{diffusion coefficient} $g(t)\D{{w}}$ follows Ito integration \cite{kara_and_shreve}. 
The diffusion coefficient $g$ regulates the amount of Gaussian noise that is added to the process, and the drift $f$ mainly affects the mean of $X_t$ (see \cite[(6.10)]{kara_and_shreve}) in the case of linear SDEs. The process state $X_t$ follows a Gaussian distribution \cite[Ch. 5]{sarkkaAppliedStochasticDifferential2019}, called the \emph{perturbation kernel}:
\begin{equation}
\label{eq:perturbation-kernel}
    p_{0t}( X_t| X_0,  Y) = \mathcal{N}_\mathbb{C}\left( X_t; \mu_t(X_0, Y), \sigma_t^2 {I}\right).
\end{equation}
We call $\mu_t(X_0, Y)$ the \textit{mean evolution} and $ \sigma_t$ the \textit{variance evolution} as they describe how the mean and variance of the process state $X_t$ are evolving over the diffusion time $t$. If we can find analytically closed-form solutions for the mean and variance evolution, then \eqref{eq:perturbation-kernel} allows us to efficiently compute the process state $X_t$ for each $t$ by calculating
\begin{equation} \label{eq:eff_sampling1}
    X_t = \mu_t(X_0, Y) + \sigma_t z,
\end{equation}
with $z \sim \mathcal N_{\mathbb{C}}(0,1)$. By Anderson \cite{anderson1982reverse}, each forward SDE as in \eqref{eq:fsde} can be associated to a reverse SDE:
\begin{equation}\label{eq:plug-in-reverse-sde}
    \D{ X_t} =
        \left[
            - f( X_t,  Y) + g(t)^2  \nabla_{ X_t} \log p_t( X_t| Y)
        \right] \D{t}
        + g(t)\D{\bar{ w}}\,,
\end{equation}
where
$\D{\bar{ w}}$ is a Wiener process going backwards in time. In particular, the reverse process starts at $t=T$ and ends at $t=0$. Here $T < T_{\text{max}}$ is a parameter that needs to be set for practical reasons as the last diffusion time-step $T_{\text{max}}$ is only reached in limit.
The \emph{score function} $\nabla_{ X_t} \log p_t( X_t| Y)$ is approximated by a NN called \emph{score model} $s_\theta( X_t,  Y, t)$, which is parameterized by a set of parameters $\theta$.
Assuming that $s_\theta$ is available, we can generate an estimate of the clean speech $X_0$ from $Y$ by solving the reverse SDE \eqref{eq:plug-in-reverse-sde}.

\subsection{Latency considerations for streaming data} \label{sec:utterance_based}
In contrast to the offline scenario, where an entire signal is available and can be operated on, the streaming scenario is more challenging: the input signal arrives one frame at a time and the output must be produced within a predefined amount of a time to prevent adding further latency to the system. In our case, we must process a chunk of overlapping frames faster than the hop length's time. Application of score-based diffusion models such as \cite{lay202interspeech, journal} on streaming data requires solving the reverse process within the time constraints defined above. Since the reverse \ac{sde} calls the score model several times, current score-based models such as \cite{lay202interspeech, welkerinter2022} are not usable for streaming data as even a single call of the score model exceeds the hop length as chosen in \cite{lay202interspeech, welkerinter2022}.

\section{Proposed method: Diffusion Buffer}
\begin{figure}[t]
  \centering
  \includegraphics[scale=1.0]{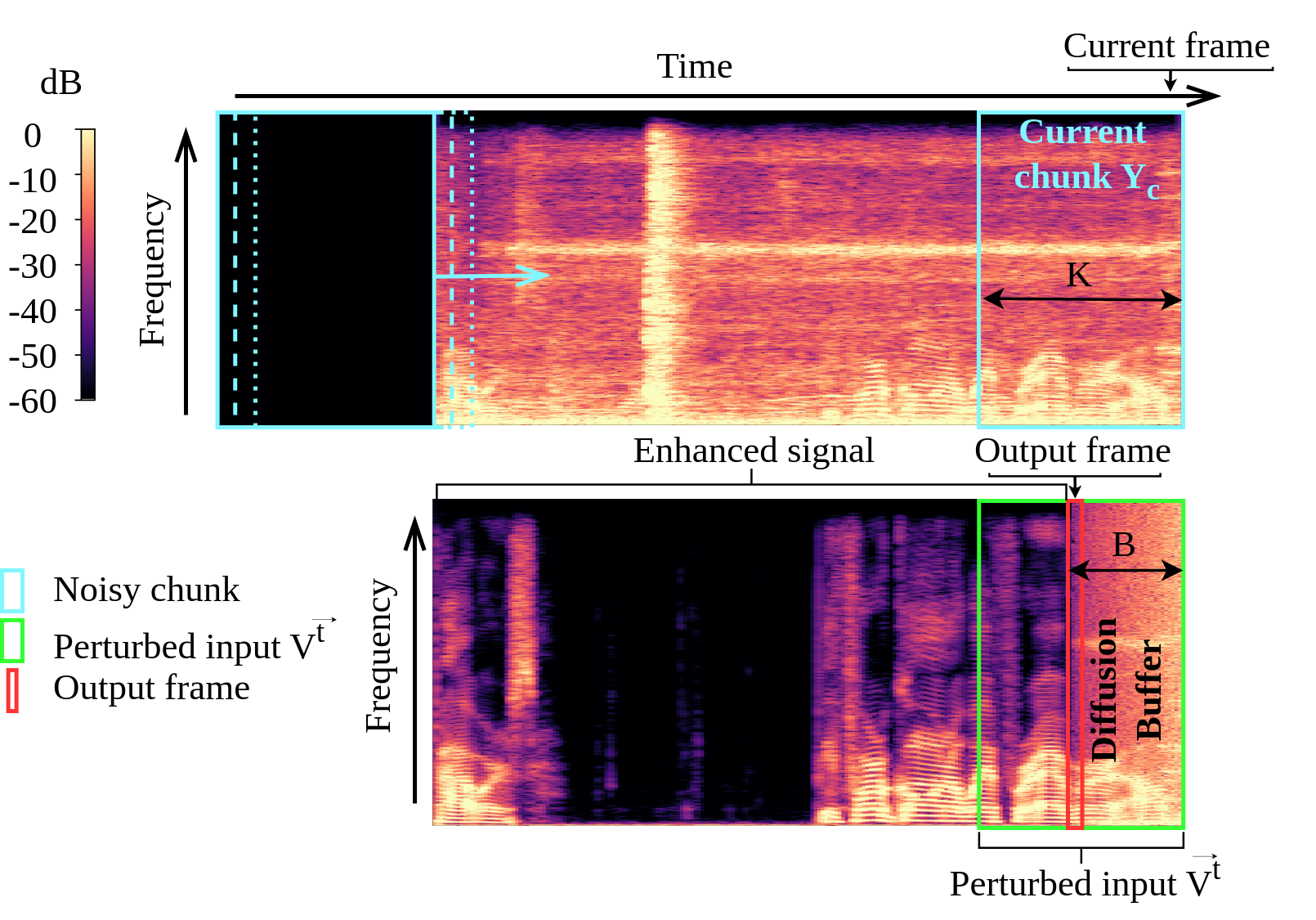}
  \caption{The \ac{db} scheme for streamed data. The top part shows the noisy stream $Y_s$. The bottom presents the enhanced signal. In $\Vt$, we see exponentially increasing noise in the \ac{db}. The output frame is indicated in red at the beginning of the \ac{db}.}
  \label{fig:diffgate}
\end{figure}

Inspired by \cite{rollingdiff, fifo}, we propose to align the diffusion time-steps with the time axis of the noisy mixture. To this end, we introduce a Diffusion Buffer containing the last $B$ frames, whereas the current frame $\mathbf R \in \mathbb{C}^{F \times 1}$ is placed at the end of this buffer and past frames are closer to the beginning of the buffer.
Within this buffer, frames that are closer to the end are modeled to be at larger diffusion time-steps and therefore contain more noise. Equivalently, frames that are further in the past are progressively denoised until they become fully denoised. Consequently, with each new frame added into the buffer, we output a denoised frame that lies further in the past. This is depicted in Fig.~\ref{fig:diffgate}. Since there is a considerable delay between the current frame $R$ and the output frame, this approach increases algorithmic latency. The amount of added latency can be adjusted by the choice of $B$, allowing for a trade-off between added latency and enhancement performance, as will be shown below.
An important advantage of this approach is that within the time of each hop length, the reverse process only requires a single evaluation of the score. This is crucial, as multiple calls to the score model are computationally infeasible for online processing.

Mathematically, we formulate this process as follows: we fix $B \leq K$ as the number of reverse steps taken to enhance the frames of the noisy mixture.
Let $\seqt = (t_1, \dots, t_B)$ be an ascending sequence of diffusion time-steps, i.e $0 < t_1 < \dots < t_B = T_{\text{max}}$. Then we call $\Vt \in \mathbb{C}^{F \times K}$ the \textit{perturbed input} and define it as

\begin{equation} \label{eq:V}
\Vtfk \coloneqq
\begin{dcases*}
S(l,k)
   & if  $k < K-B$\,, \\[1ex]
X_{t_{g(k)}}(l, k)
   & if $k \geq K-B$\,.
\end{dcases*}
\end{equation}
where $g(k) = k - (K-B) + 1$ and $X_{t_{g(k)}}(l, k)$ can be computed via \eqref{eq:eff_sampling1}. In this formulation, the current frame from streamed data is placed at the last frame $V(l,K)^{\seqt}$ which contains the most amount of Gaussian noise injected by the diffusion process. Moreover, we see that the frames $\Vtfk$ for $K-B\leq k \leq K$ are becoming clean when close to $K-B$ and noisy when $k$ is close to $K$. All other frames outside of this buffer are already assumed to be clean. The collection of frames $\Vtfk$ with $K-B\leq k \leq K$ can also be thought of as a buffer on which we perform the diffusion process. Hence, we call it the \textit{Diffusion Buffer}.

\subsection{Training} \label{sec:training}
\vspace{-4px}
For training, we fix the number of frames $K$, frequency bands $F$, the number of reverse steps $B$ in the Diffusion Buffer, and the smallest diffusion time-step $\epsilon > 0$.
We first sample a pair of clean and noisy files from the dataset. Then we pad the clean and noisy files with $K-1$ leading zeros in the STFT domain to mimic initialization when processing streamed data. In addition, we randomly crop a segment of $K$ frames from the clean and noisy file which we call $S=X_0$ and $Y$ respectively for the sequel. 
Second, we uniformly randomly sample an ascending sequence $\seqt = (t_1, \dots, t_B)$ of Diffusion time-steps with $t_1 = \epsilon > 0$. Third, based on \eqref{eq:eff_sampling1}, we compute for $K-B \leq k \leq K$
\begin{equation} \label{eq:eff_sampling2}
    X_{t_{g(k)}}(l, k) = \mu_{t_{g(k)}}(X_0(l,k), Y(l,k)) +\sigma_{t_{g(k)}} z(l,k)
\end{equation}
with $z(l,k) \sim \mathcal N_{\mathbb{C}}( 0, 1)$. We arrange $z(l,k)$ in a matrix as $\mathbf Z \in \mathbb{C}^{F\times B}$, likewise $\mathbf \Sigma \in \mathbb{C}^{F\times B}$. Fourth, we use \eqref{eq:eff_sampling2} to calculate $\Vtfk$ as in \eqref{eq:V} for all frequencies $1\leq f \leq F$ and frames $1\leq k \leq K$. Last, we optimize on the denoising score matching loss:

\begin{equation}\label{eq:training-loss}
      \argmin_\theta \mathbb{E}_{t,(\mathbf X_0,\mathbf Y), \mathbf Z, \mathbf \Vt|(\mathbf X_0,\mathbf Y)} \left[
        \norm{\mathbf s_\theta(\mathbf \Vt, \mathbf Y, t) + \frac{\mathbf Z}{\mathbf \Sigma}}_2^2
    \right]
\end{equation}
where the division of matrices $\frac{\mathbf Z}{\mathbf \Sigma}$ is meant to be elementwise. Note that for the score model input we have $\mathbf \Vt, \mathbf Y \in \mathbb{C}^{F\times K}$, but the score model's output is in $\mathbb{C}^{F \times B}$.

\begin{algorithm}
\caption{Proposed Online Speech Enhancement}\label{alg:online2}
\begin{algorithmic}[1]
\Require Reverse process $u_\theta$ with trained score model $s_\theta$, noisy stream $\mathbf Y_s$, chunk size $K$, fixed diffusion time-steps $\seqt = (t_1, \dots, t_B)$
\State $\hat{\mathbf S}$ $\gets$ [] \Comment{initialize output}
\State $\Vt$ $\gets [0, \dots, 0]$ \Comment{$K$ empty frames}
\For{frame $\mathbf R$ in $Y_s$}
    \State $\mathbf Y_c \gets$ last $K$ received frames \Comment{initialize with 0}
    \State $\Vt$ pops  \Comment{removes its first frame}
    \State $\Vt$ appends $ \mathbf R + \sigma_{t_B}\mathbf  Z$ \Comment{$\mathbf Z\sim \mathcal N_{\mathbb{C}}(\mathbf 0, \mathbf I_{F \times 1})$}
    \State $\Vt$ $\gets u_\theta(\Vt, \mathbf Y_c, \seqt)$   \Comment{only one $s_\theta$ call}
    \State $\hat{\mathbf S}$ appends $B$-th last frame of $\Vt$
\EndFor
\State \textbf{Return} $\Hat{S}$
\end{algorithmic}
\end{algorithm}

\vspace{-0.5cm}
\subsection{Online inference for streamed data}
Once we have a trained model as described in Section \ref{sec:training}, we run inference as described in Algorithm~\ref{alg:online2} and illustrated in Fig.~\ref{fig:diffgate}. Let $\mathbf Y_s$ be an infinite stream of data in the STFT domain and assume we receive the frame $\mathbf R$ in the for-loop of Algorithm~\ref{alg:online2}. We then add in line 6 an amount of Gaussian noise so that the random variable $\mathbf R' = \mathbf R + \sigma_{t_B} \mathbf Z$ follows the perturbation kernel \eqref{eq:perturbation-kernel}, meaning $\mathbf R'$ is at diffusion time-step $t_B$. As usual, we then run one reverse step for $\mathbf R'$ where the score function is approximated by the trained score model. Consequently, $\mathbf R'$ is now at diffusion time-step $t_{B-1}$. In fact, we run the reverse step for all frames within the \ac{db}. In particular, this means that the $B$-th last frame which was before the reverse step at diffusion time-step $t_1 = \epsilon$ is now at diffusion time-step $0$. We therefore have enhanced this frame and output it to the listener. The output frame is also depicted in Fig.~\ref{fig:diffgate} which is at the beginning of the \ac{db}.
Hence, the latency of this approach is given by $h_s \cdot B$, where $h_s$ is the hop length in seconds, plus the compute time for one output frame, which should be smaller than the hop-length $h_s$ for online processing.
Note that the reverse process only enhances frames within the \ac{db}, all other frames are not processed as they are already enhanced (see also Fig.~\ref{fig:diffgate}). As we slide over the streamed data as shown in Fig.~\ref{fig:diffgate}, we ensure that each frame has undergone $B$ reverse steps before it is enhanced. The zero part on the left of $\mathbf Y_s$ in Fig.~\ref{fig:diffgate} is required for initialization. For this, we intentionally padded the training data with leading zeros in the first step in Section \ref{sec:training} to match the initialization.

\section{Experimental Setup}
\subsection{Data representation}  
\vspace{-5px}
\label{sec:exp:data}
Each audio input, sampled at 16 kHz, is converted to a complex-valued \ac{stft}. As in \cite{journal}, we use a window size of 510 samples (32 ms), a hop length of 256 samples ($h_s=16$ms), and a periodic Hann window. The input to the score model is cropped randomly to $K=128$ time frames, resulting in approximately 2 seconds of data (see Section \ref{sec:training}). A magnitude compression is used to compensate for the typically heavy-tailed distribution of \ac{stft} speech magnitudes~\cite{gerkmann2010empirical}. Each complex coefficient $v$ of the \ac{stft} representation is transformed as $\beta |v|^\alpha \mathrm e^{i \angle(v)}$ with $\beta=0.15$ and $\alpha=0.5$, as in \cite{journal, lay202interspeech}.

\subsection{Score model} \label{sec:exp:score model}
\vspace{-5px}
For the score model $s_\theta(\mathbf X_t, \mathbf Y, t)$, we employ the Noise Conditional Score Network (NCSN++) architecture. The original NCSN++ architecture used in \cite{journal, song2021sde} had 65M parameters. We reduced the network capacity by adapting its channel dimension from $128$ to $96$, reducing the number of Downsampling/Upsampling blocks from 6 to 4, and decreasing the number of residual blocks from 2 to 1. The resulting network has only 18M parameters.

For the proposed \ac{db} method we need to adapt the network in two ways. First, the time-embeddings $t$ originally were only added to the channel dimension of the features of NCSN++. In our case, we now have a sequence of diffusion time-steps that need to be adapted to the channel and frame dimension of the features of the network. To this end, we use a Conv2D layer with stride to match the channel and frame dimension to the feature. This resulted in an additional 300.000 parameters. Second, the network output was originally of the same shape as its input $\mathbf Y$. Since we train on the loss from \eqref{eq:training-loss}, we need to crop the output of the NN to contain only the last $B$ frames, instead of taking all $K$ frames.

We train the network on the loss described in Section \ref{sec:training}. The optimizer we use is the ADAM optimizer \cite{kingma2015adam} with a learning rate of $10^{-4}$ and a batch size of 32. For smoothing the network parameters along the training epochs, we employ an exponential moving average of the score model's parameters \cite{journal,song2021sde} (decay of 0.999). We trained for 250 epochs.

\begin{figure}[t!]
\vspace{-1.0cm}\includegraphics[width=0.9\linewidth]{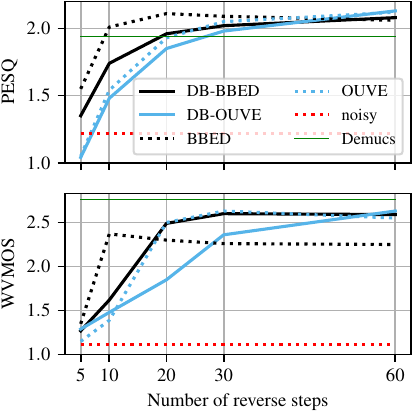}
    \caption{Comparing OUVE, BBED against their DB versions on the filtered EARS-WHAM test set. We experimented with different number of reverse steps. This is also the buffer size $B$ for the \ac{db} versions with algorithmic latency equal to $h_s \cdot B$.}\label{fig:pesq}
\end{figure}

\subsection{Datasets and metrics} \label{sec:exp:dataset}
\vspace{-5px}
We use the publicly available dataset EARS-WHAM \cite{richter2024ears}, originally recorded in 48 kHz. We downsampled the data to 16 kHz and filter it by removing files of type ``whisper'', ``highpitch'' and ``lowpitch''. We also used only pairs of clean and noisy with a Signal-to-Noise ratio in [0, 15] db. This dataset has 54 hours for training, 1.1 hours of validation, and 2 hours for testing. \\
\indent We evaluate the performance on the perceptual metric wideband PESQ \cite{rixPerceptualEvaluationSpeech2001}. We also evaluate on a reference-free metric WVMOS \cite{wvmos} using a NN to predict MOS values. Note that this metric sometimes produces negative MOS values. In these cases, we simply set the WVMOS value to 1.0. Moreover, the RTF is defined as the processing time for one iteration in Algorithm \ref{alg:online2} divided by $h_s$.

\subsection{SDEs and baselines}\label{sec:exp:baseline}
\vspace{-0.2cm}
We will compare against OUVE \cite{journal} and BBED \cite{lay202interspeech} with the data representation from Section \ref{sec:exp:data}. We use the parameterization as described in \cite{lay2024analysis}. Specifically, the drift term of OUVE is $f(\mathbf X_t,\mathbf Y) = \gamma(\mathbf Y- \mathbf X_t)$, and BBED's drift term is $\frac{\mathbf Y-\mathbf X_t}{1-t}$. Moreover, both SDEs use the diffusion term $g(t) = ck^t$. We refer to \cite{lay2024analysis} for the closed-form solution for the mean and variance evolution, which is used to compute \eqref{eq:eff_sampling1} or \eqref{eq:eff_sampling2}.
We select the parameters $c,k,\gamma$ as in \cite{lay2024analysis}. Precisely, for OUVE we have $\gamma=1.5$, $c=0.01, k=10$ with a reverse starting point of $1.0$. For BBED we select parameters $c=0.08$, $k=2.6$  with a reverse starting point of $0.8$. For inference, we do not use an online streaming framework but rather enhance noisy files by utterance-based processing as it is done in \cite{journal, lay202interspeech}. We use NCSN++ as described in Section \ref{sec:exp:score model} with the reduced capacity. 

The proposed method \ac{db} is trained with the same parameterizations of BBED and OUVE which we call DB-BBED and DB-OUVE respectively. Additionally, we employ the NCSN++ parameterization used for the baselines. We also incorporate the necessary changes outlined in Section \ref{sec:exp:score model}.

We also compare against the real-time capable discriminative network DEMUCS \cite{defossez2020real} operating in the time-domain.

\section{Results}
\vspace{-2px}
We train the proposed \ac{db}-OUVE and \ac{db}-BBED on the filtered 16 kHz EARS-WHAM dataset and experimented with the buffer length $B = 5,10,20,30,60$.

First, we discuss, as in Section \ref{sec:utterance_based} why the vanilla diffusion models OUVE and BBED do not operate in real-time on streamable data as their RTFs $\gg 1$. Compared to their original implementation in \cite{journal, lay202interspeech} where the network has 65M parameters, we use in this work a reduced version of NCSN++. We intentionally reduced the NCSN++ architecture as described in Section \ref{sec:exp:score model} to ensure that the network's processing time on a laptop with an NVIDIA RTX 4080 is smaller than the duration of one hop length of $h_s<16$ ms on chunks with $K=128$ frames. In fact, the processing time of taking one reverse step with the reduced NCSN++ is $\approx 14$ ms (RTF is $\frac{14}{16}$). When now applying BBED or OUVE to streamable data, during each hop length, the reduced NCSN++ is called $N$ times where $N$ is the number of reverse steps (RTF $= N\frac{14}{16}$). This approach becomes computationally infeasible for $N>2$ and will not operate in real-time on the given hardware. In addition for $N=1$, we report results that OUVE and BBED are even much worse than the noisy input. In contrast, the versions with a DB call the score model only once during the iteration in Algorithm \ref{alg:online2} at the cost of increasing the latency by $h_s \cdot B$.

Second, we observe in Fig.~\ref{fig:pesq} that the baselines OUVE and BBED slightly outperform their \ac{db} versions in terms of PESQ for $5,10,20,30$ numbers of reverse steps. However, for 60 reverse steps, the proposed DB-BBED and DB-OUVE methods are marginally outperforming their baseline counterparts by 0.05 PESQ. For WVMOS the proposed DB methods are at least on par with the baselines for all reverse steps (except for OUVE and DB-OUVE at 5 reverse steps). DB-BBED even outperforms BBED for 20, 30, 60 reverse steps. In addition, we see from Fig.~\ref{fig:pesq} that with already 20 reverse steps we achieve reasonable results. Hence, the proposed method performs well with an algorithmic latency of 320 - 960 ms. 

Third, we observe that the proposed methods are outperformed by DEMUCS in WVMOS, but achieve higher PESQ values when $N>30$. We report results that the proposed methods ($N>30$) are as intelligible as DEMUCS, as their ESTOI \cite{jensen2016algorithm} values differ only by 0.01.

The SDE parameterization of BBED has a larger variance schedule compared to OUVE, as shown in \cite{lay2024analysis}. A higher variance schedule reduces the number of reverse steps needed for enhancement, which is why BBED outperforms OUVE in terms of PESQ and WVMOS when fewer reverse steps are used (see dotted lines in Fig.~\ref{fig:pesq}). This advantage extends to score-based diffusion models with the proposed \ac{db}, where \ac{db}-BBED also outperforms \ac{db}-OUVE for a few number of reverse steps (see solid lines in Fig.~\ref{fig:pesq}). Better performance with fewer reverse steps is crucial as it reduces algorithmic latency.

\section{Conclusion}
In this work, we successfully adapted score-based diffusion models to process streamed audio data online. Inspired by \cite{rollingdiff, fifo}, we denoise noisy frames through physical time. Unlike standard diffusion models that do not take the physical time axis into consideration, we introduce the Diffusion Buffer, where the position of each frame is taken into account and frames further from the present are progressively enhanced. This approach is computationally feasible as the score model is evaluated only once while waiting for new noisy data. The proposed Diffusion Buffer enables a trade-off between SE performance and latency, given by the number of reverse steps. At latencies between 320 and 960 ms, the proposed methods are running online on a consumer GPU achieving SE performance comparable with existing offline diffusion-based methods.

\section{Acknowledgments}
Funded by the Deutsche Forschungsgemeinschaft (DFG, German Research Foundation) -- 545210893, 498394658.

Funded by the Federal Ministry for Economic Affairs and Climate Action (Bundesministerium für Wirtschaft und Klimaschutz), Zentrales Innovationsprogramm Mittelstand (ZIM), Germany, within the project FKZ KK5528802VW4.

The authors gratefully acknowledge the scientific support and HPC resources provided by the Erlangen National High Performance Computing Center (NHR@FAU) of the Friedrich-Alexander-Universität Erlangen-Nürnberg (FAU) under the NHR project fac101ac. NHR funding is provided by federal and Bavarian state authorities. NHR@FAU hardware is partially funded by the German Research Foundation (DFG) – 440719683.

\bibliographystyle{IEEEtran}
\bibliography{ref}

\end{document}